ORIGINAL PAPER

# A comparative study on viscoelastic properties of polymeric composites measured by a longitudinal free vibration non-destructive test and dynamic mechanical thermal analysis

Mohammad Mehdi Jalili · Amir Soheil Pirayeshfar · Seyyed Yahya Mousavi

**Abstract** In this study, viscoelastic properties of polymeric composites were investigated through a non-destructive test (NDT) method based on longitudinal free vibration. First, three different polyester composites reinforced separately with carbon, glass, and hemp fibers, as well as, one polyester composite sample reinforced with poplar wood powder were manufactured via pultrusion and hand lay-up methods, respectively. In the proposed resonant free vibration non-destructive method, each rod-shaped test specimen was hit by a hammer at one end of specimen and immediately the acoustic response of the specimen was recorded by a microphone at the other end of the sample. Subsequently, the recorded sounds were analysed using fast Fourier transform technique. Then, frequency of the first mode of vibration for each composite specimen was utilised to calculate modulus of elasticity. Further, decrement in vibration energy as a function of time was examined to obtain loss parameter (tan δ) of the provided composites. Moreover, parameters (i.e., elastic modulus and tan δ) were also compared with those determined by dynamic mechanical thermal analysis (DMTA). It is found that the results obtained from the examined non-destructive test method are in consistent with those measured by DMTA approach, providing reliable, accurate and quick responses.

**Keywords** Non-destructive test · Longitudinal free vibration · Elastic modulus · Loss parameter · Polyester composites

M. M. Jalili (✉) · A. S. Pirayeshfar · S. Y. Mousavi
Department of Polymer Engineering, Science and Research Branch, Islamic Azad University, Tehran, Iran
e-mail: m.jalili@srbiau.ac.ir

## Introduction

Static and dynamic destructive tests such as tensile and dynamic mechanical analysis (DMA) are extensively employed to measure viscoelastic properties of polymeric materials. Although these methods are so practical, they also possess some problems. For example, test procedure is irreversible, because the sample is destroyed; achievement of desirable sample geometry according to the standards is relatively difficult and also long procedure time is usually required. Moreover, investigating the polymer changes and polymer degradation for one test specimen during exposure time is impossible [1–3].

In contrast, non-destructive test (NDT) methods causing no damage and undesirable effects on the samples, can accurately estimate viscoelastic properties of materials in a short time. Recently, the non-destructive evaluation of elastic properties of composite structures is more commonly used [4, 5]. NDTs are categorised methodologically [6]. One of them is called resonant vibration testing which is involved mechanically vibrating of a test specimen in either of torsional, transverse, or longitudinal vibration modes over a range of frequencies [6, 7]. This method has been utilised by some authors to investigate and analyse the properties of ceramic [8], concrete [9] and especially wood [10, 11]. Schmidt et al. measured Young's modulus of molding compounds and examined the effectiveness of temperature on compounds' modulus with resonance method. They also reported a good agreement between NDT results and mechanical measurements [12, 13]. In spite of the fact that it is a powerful, precise and quick method to determine viscoelastic properties of materials, there is a lack of studies in the field of polymers and composites. Although some attempts have been made in this field, yet, this method has not been well established.

In this work, we use a resonant vibration NDT method based on longitudinal free-vibration mode to examine viscoelastic properties of fiber and powder reinforced composites. We also compare the NDT results with those measured by DMTA approach.

### Experimental

Materials

Isophthalic unsaturated polyester resin, Boshpol 751129, was purchased from Bushehr Chemical Industries Co. (Iran). Amounts of 1 vol% methyl ethyl ketene peroxide (Pamookaleh Co., Iran) and 0.9 vol% cobalt octoate (Saveh Chemicals Co., Iran) were also added to the polyester resin according to the data sheet recommendations as initiator and accelerator, respectively. Carbon fiber (T300) from Troyca Co. (USA), glass fiber (WR3) from Camelyaf Co. (Turkey) and hemp fiber (Industrial grade) from Iran Kenaf Co. (Iran) were provided and used as fiber reinforcements, separately. In addition, poplar wood powder with mesh size of 30–40 was utilised as powder reinforcement. Young's moduli of the carbon, glass, and hemp fibers were 230, 60, and 30 GPa, respectively. In addition, Young's modulus of the polyester resin was measured by tensile tester (model Elima, Iran) according to ASTM D638 and the average value for three replicates was recorded as 3 GPa.

Procedure of sample preparation

Pultrusion and hand lay-up methods were used to prepare the specimens. Fiber composites, rod-shaped specimens with diameter of 9 mm were produced through pultrusion method and then were cut with length of 60 cm.

In addition, hand lay-up method was used to prepare rectangular test specimens of wood powder composites with dimensions of (2.5 × 2.5 × 20 cm). To produce these samples, first, wood powder was mixed with polyester resin at a specific ratio, and then, the mixture was used in hand lay-up procedure. Each specimen cured in ambient temperature and then tested after 1-month exposure in ambient condition. Table 1 shows compositions and selected labels of the prepared specimens. Note that the shape of cross-section of test specimens does not influence the calculations employed in this method.

Dynamic mechanical thermal analysis

Dynamic mechanical thermal analysis (DMTA) was carried out to examine the viscoelastic properties of the prepared samples using DMTA-Triton model Tritec2000 DMA (UK).

Longitudinal free vibration NDT

To implement this technique, the set-up as shown in Fig. 1 was prepared. First, each test specimen was hold from its center and was hit by a wooden hammer at one end of specimen. To analyse the acoustic response of the specimen, a microphone was positioned in the other side of sample. Subsequently, the respond vibrating sound was recorded by Audacity software as a wave-format file. Finally, the recorded sounds were analysed. All the measurements were carried out in the ambient temperature (23 ± 2 °C).

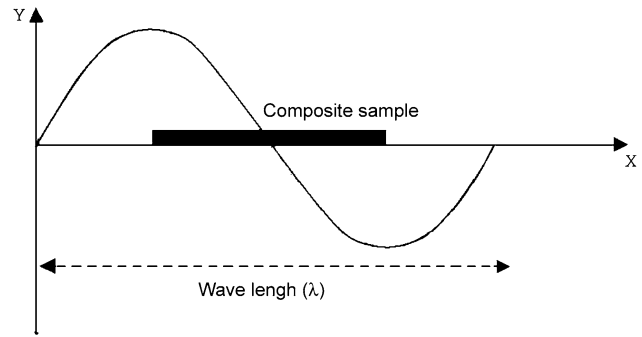

Fig. 2 Position of node and anti-nodes in a specimen hit by a wooden hammer at one of its ends

Table 1 Label and composition of samples

| Sample label | Components |
| --- | --- |
| PsCf | 70 wt% carbon fiber + 30 wt% polyester resin |
| PsGf | 70 wt% glass fiber + 30 wt% polyester resin |
| PsHf | 70 wt % hemp fiber + 30 wt% polyester resin |
| PsPw | 50 wt % poplar wood powder + 50 wt% polyester resin |

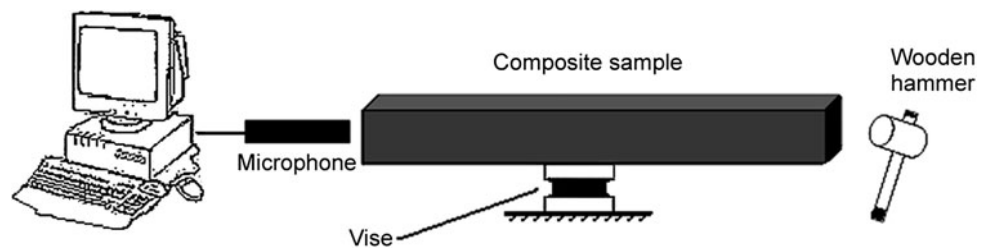

Fig. 1 Set-up of longitudinal free vibration non-destructive test

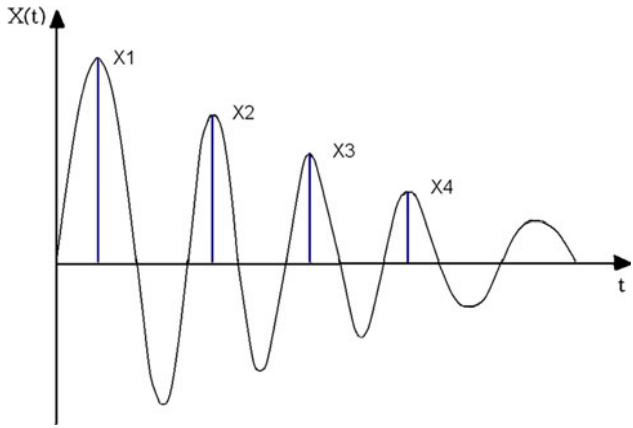

**Fig. 3** Schematic view of amplitude decrement of the first mode of vibration with time

Theory

Longitudinal elastic modulus could be determined by the aid of ultrasonic velocity according to Eq. (1) [14–16]:

$$E = \rho V^2 \varphi(\sigma) \quad (1)$$

$$\varphi(\sigma) = \frac{[(1+\sigma)(1-2\sigma)]}{(1-\sigma)} \quad (2)$$

where, $E$ is elastic modulus, $\rho$ specific density, $V$ ultrasonic velocity, and $\sigma$ is Poisson's ratio.

The range of $\sigma$ for polymeric composite materials has been mostly reported from 0.25 to 0.3; therefore, $\varphi(\sigma)$ could be approximately replaced by 0.8 [17–20]

In general, ultrasonic velocity in a specimen could be determined by Eq. (3) [11]:

$$V = f \times \lambda \quad (3)$$

where, $V$ is ultrasonic velocity in a specimen, $f$ is resonance frequency of the first mode of vibration when the vibration resonance occurs in the specimen for the first time. In addition, $\lambda$ is wave length and can be calculated from Eq. (4) [11]:

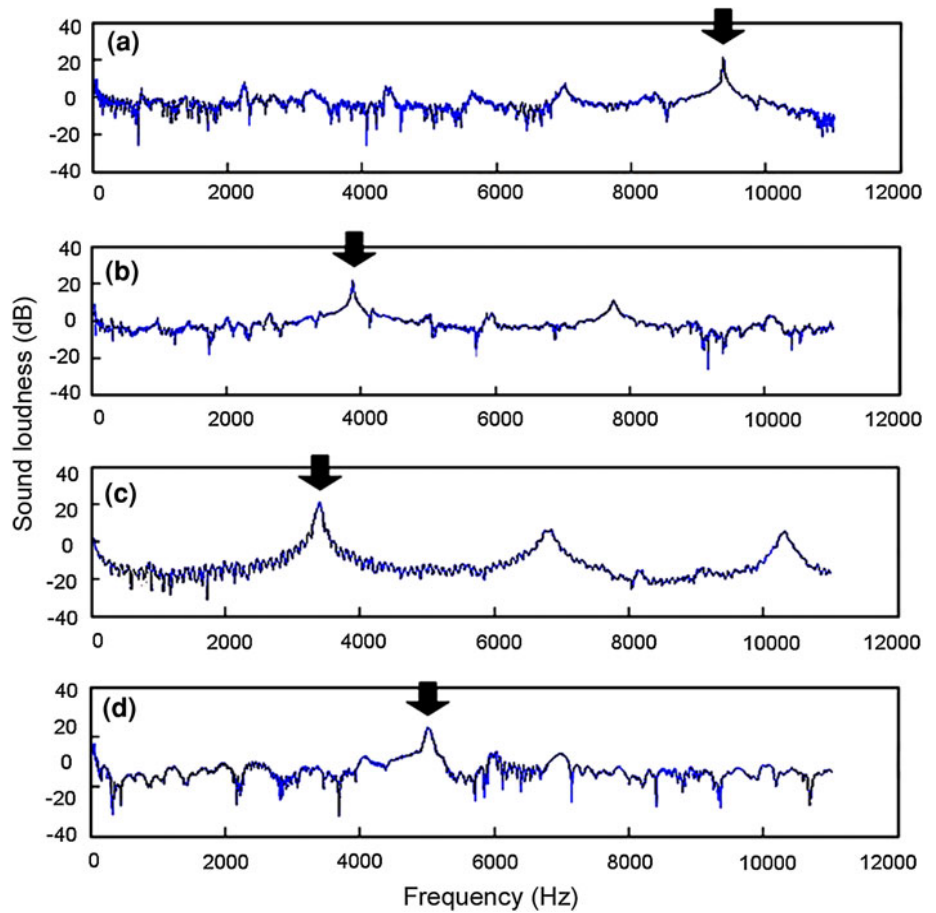

**Fig. 4** Analysis of the recorded sound using FFT technique for **a** PsCf, **b** PsGf, **c** PsHf, and **d** PsPw composite specimens

**Fig. 5** First mode of vibration curves derived from Fig. 4 for **a** PsCf, **b** PsGf, **c** PsHf, and **d** PsPw composite specimens

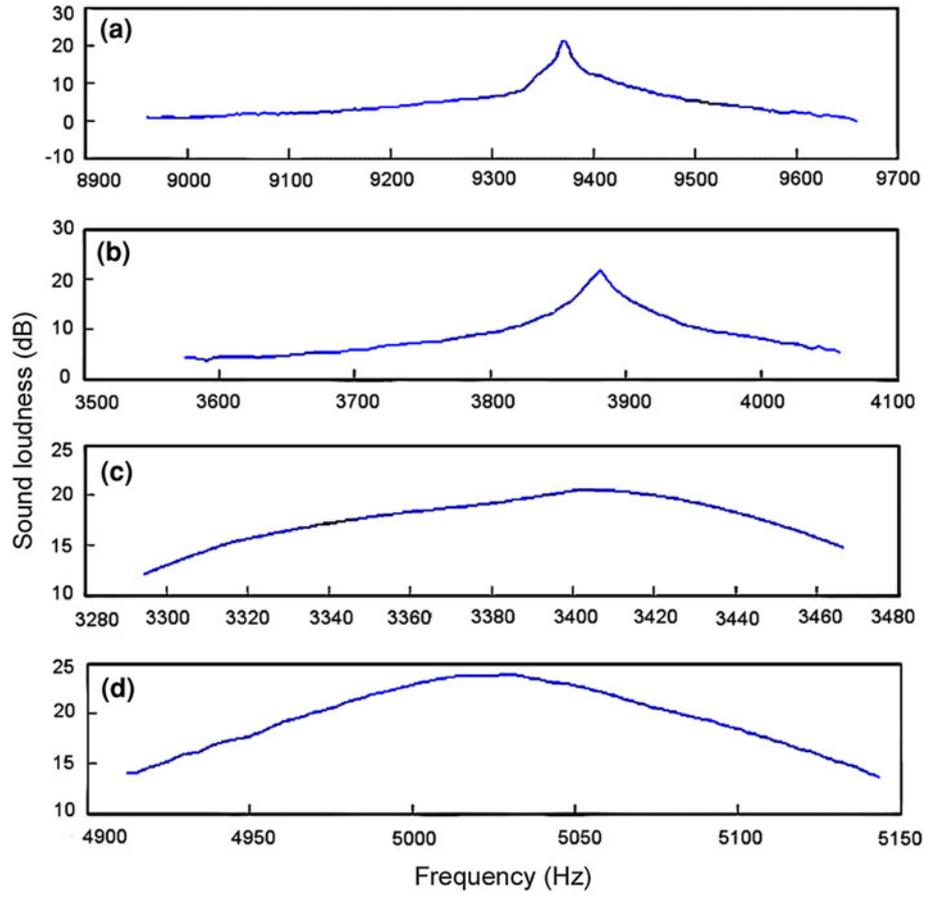

$$\lambda = \frac{2L}{n} \qquad (4)$$

where, $L$ is the length of specimen and $n$ is the number of resonance mode. Note that each vibrating material possesses infinitive modes of resonance. For the first mode of vibration, $n$ is equal to 1, therefore, the wave length of first mode of vibration can be obtained as follows:

$$\lambda = 2L \qquad (5)$$

In fact, there is a node on the vibration wave at the centre of specimen combined with two anti-nodes at the both ends of the specimen as shown in Fig. 2. According to the positions of node and two anti-nodes corresponded to the first mode of vibration, the wave length is equal to twice the length of specimen [21].

After each impulse, as described earlier, the sample starts vibrating. But, polymeric composites are not perfect elastics, therefore, the vibration energy is reduced after a while due to the internal friction and cohesion between molecules. This attenuation occurs due to material damping which is affected by the type of material. Therefore, decrement in vibration energy as a function of time could be a relevant key factor to determine the loss parameter (tan δ) of materials, as expressed in Eq. (6) [10]:

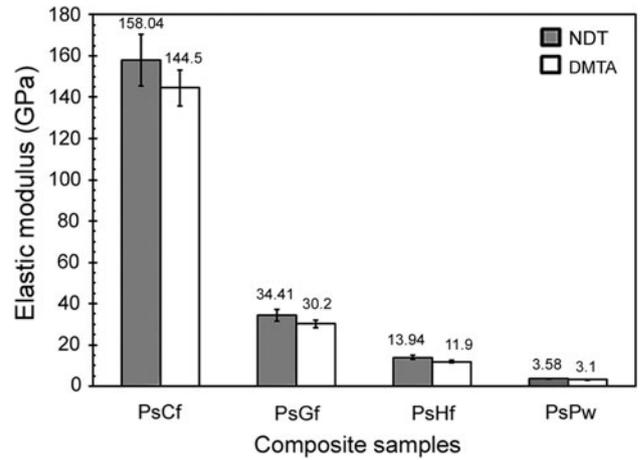

**Fig. 6** Elastic moduli obtained from NDT and DMTA methods for PsCf, PsGf, PsHf and PsPw composite specimens

$$\tan\delta = \frac{\lambda}{\pi} \qquad (6)$$

where $\lambda$ is logarithmic vibrating decrement factor which could be calculated by Eq. (7) [10]:

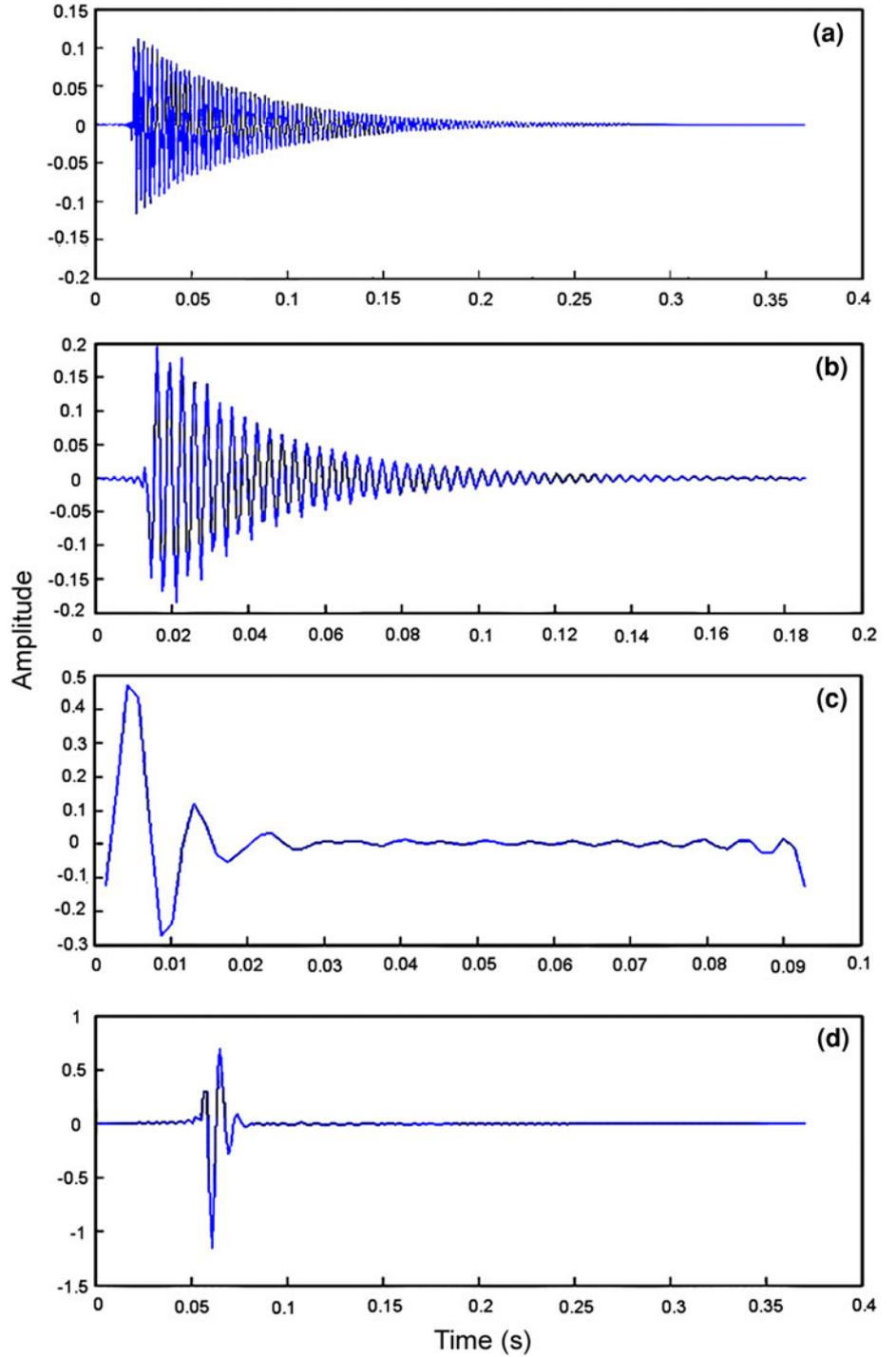

**Fig. 7** Vibration amplitude decrement of the first mode of vibration versus time for **a** PsCf, **b** PsGf, **c** PsHf, and **d** PsPw composite specimens

$$\lambda = \left(\frac{1}{n}\right) \ln \left| \left(\frac{x_n}{x_{n+1}}\right) \right| \qquad (7)$$

where, $n$ is time parameter, $x_1$ and $x_{n+1}$ are the first and $(n + 1)$th amplitudes of vibration, respectively (Fig. 3). In addition, Eq. (8) could be simply achieved by the rearrangement of Eq. (7) as follows:

$$\ln x_{n+1} = -\lambda n + \ln x_1 \qquad (8)$$

## Results and discussion

A sound wave comprises three components: loudness, frequency and time. To display each sound in terms of its components, all the recorded sounds were analysed by the means of fast Fourier transform (FFT) technique. Results of analysis of the recorded sounds are depicted in Fig. 4 showing two components (loudness vs. frequency) for each specimen.

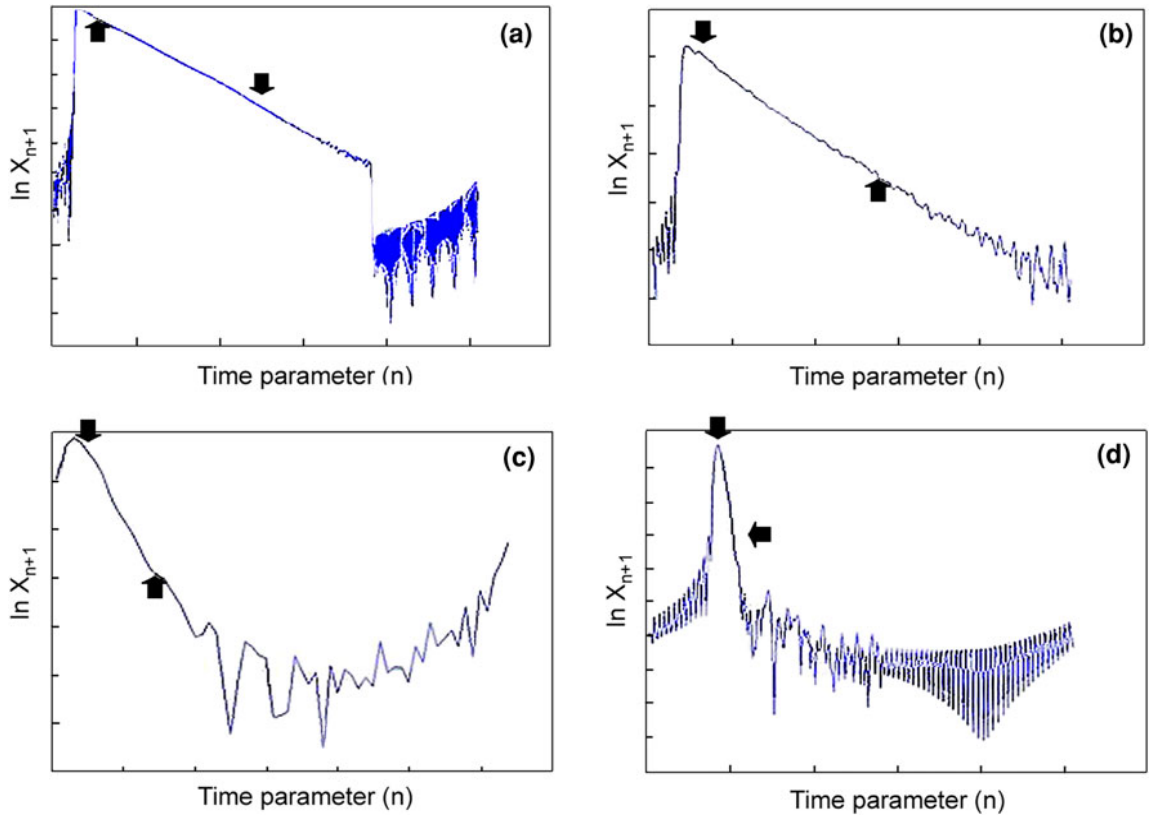

**Fig. 8** Logarithmic view of vibration amplitude decrement of the first mode of vibration versus time parameter for **a** PsCf, **b** PsGf, **c** PsHf, and **d** PsPw composite specimens

Although there are infinitive frequencies for each specimen in which vibration resonance occurs, depending on the material structure [22] it has been well documented [11, 23] that by analysing the first mode of vibration, viscoelastic properties of the specimen could be characterised. It should also be noted that other modes of vibration (e.g., second and third modes of vibration) can be utilised when a flexural free vibration NDT test is employed.

Note that the frequency related to the first mode of vibration is the prime frequency in which the vibration resonance occurs in a specimen for the first time. As in Fig. 4, the first obvious mode of vibration which occurs in specific frequency is marked in each graph. Moreover, these first modes of vibration were filtered for each test specimen. Showing these filtered modes in Fig. 5, the frequencies of the first modes can be easily read. By considering these frequency values and using Eq. (3), the sound velocity can be calculated for each sample, and then, elastic modulus can be easily determined.

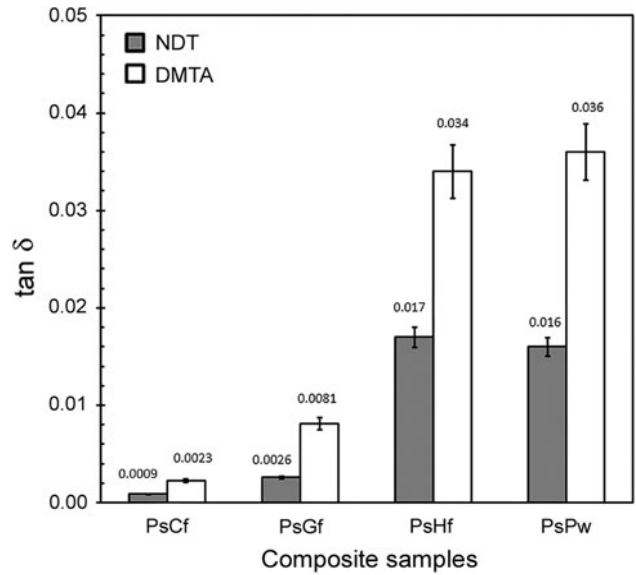

**Fig. 9** Plots of tan $\delta$ obtained by NDT and DMTA methods for PsCf, PsGf, PsHf and PsPw composite specimens

Figure 6 illustrates the obtained elastic modulus values for all composite specimens. Furthermore, the results obtained from the destructive method (i.e., DMTA) are also displayed in Fig. 6. As it is expected, the maximum modulus value belongs to the polyester composites reinforced by carbon fiber, and the modulus value of the other composites is ranked in the order of the composites reinforced by glass fiber > hemp fiber > poplar wood powder. The ranking order is identical for both NDT and DMTA methods. As in Fig. 6, there is also a very good agreement between the values of modulus obtained by NDT method and those obtained by DMTA approach measured at 25 °C. As seen, the moduli obtained from the later are lower than those calculated from the former. This may be commented by the much shorter procedure time of NDT method. When stress is applied during the DMTA experiment, creep phenomenon and phase lag of deflection with respect to stress may play a significant role in elastic modulus value of viscoelastic composites [24].

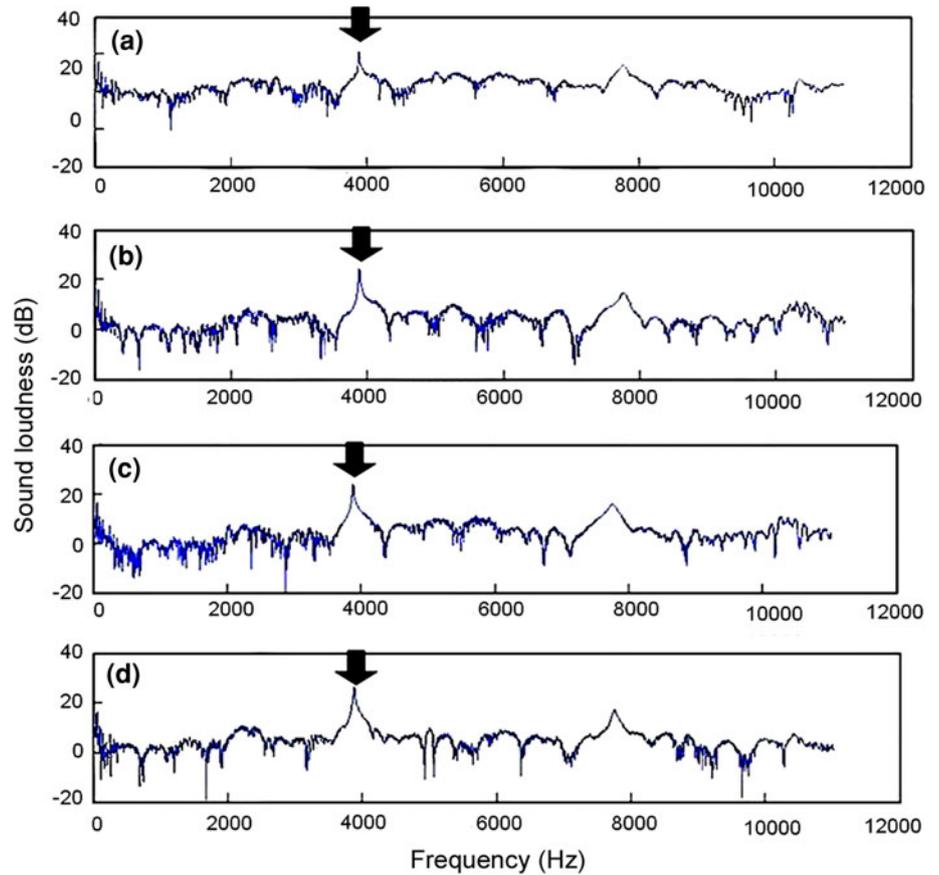

Fig. 10 Analysis of the recorded sound using FFT technique for the PsGf specimen at its four replicates

It is also noteworthy that a specific constant cannot be found to convert NDT results to DMTA ones. As shown in Fig. 6 the percentage of differences varies from 8 to 15. Such differences can be inserted in NDT values through calculation of $\varphi(\sigma)$ and/or some theoretical assumptions in this non-destructive approach. It is evident that different methods cannot exactly result in identical values due to their basic theoretical assumptions and/or practical conditions. Fortunately, a good agreement can be detected between the results measured through DMTA destructive method and those calculated from this non-destructive method, described earlier. It should be noted that in some cases based on other NDT acoustic methods, an acceptable agreement between destructive and non-destructive methods has not been discovered by other authors [25].

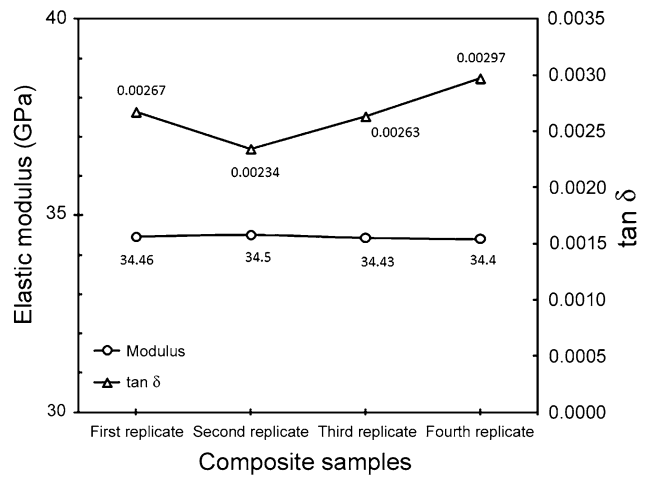

Fig. 11 Elastic modulus and tan $\delta$ plots calculated by means of NDT method from four separate replicates of PsGf specimen

Decrement of vibrating amplitude with time was also provided from the first mode of vibration (shown in Fig. 4) using FFT. Figure 7 shows variations in vibrating amplitude of test specimens versus time. As in Fig. 7, at the initial time, the amplitude vibration is very high and it then decreases with time until it reaches the zero. In other words, the vibration energy of the specimens is being damped with time; thus, it seems that the decrease in vibration energy with time could be indicative of the loss parameter (tan $\delta$) of the vibrating composite. As can be seen in the Fig. 7a, b, the amplitude decrement with time is much less for composites reinforced with carbon- and glass-fibers compared with those reinforced with hemp fiber and wood powder showing an abrupt decrease in amplitude. Such behaviour indicates that the hemp fiber- and wood powder-reinforced composites are not able to retain the vibrating energy and suddenly lose it.

According to the Eq. (8), the new curves for each test specimen are depicted in Fig. 8 (based on the data given in Fig. 7). The logarithmic decrement factor ($\lambda$) of each test specimen could be obtained from linear region of the depicted curves in Fig. 8a–d. In other words, the slope of these linear regions is equal to $\lambda$.

Finally, the loss parameter (tan $\delta$) of the examined composites were calculated by replacing logarithmic decrement factor ($\lambda$) in Eq. (7). The obtained data are presented in Fig. 9. In addition, Fig. 9 illustrates the values of tan $\delta$ obtained from DMTA. Ranking order of tan $\delta$ obtained from NDT method is approximately the same as that provided by DMTA. Nevertheless, the DMTA results reveal higher values than NDT ones. This indicates that composite samples show strong ability to damp the energy during DMTA experiments. Such behaviour may be related to the sample preparation procedure in which the composite samples were mechanically thinned to be fitted in the clamps of DMTA apparatus that may results in unsought effects on the obtained loss parameters such as induced residual stresses in the composite. Further, it may be assigned to the slipping of the rigid composite samples on the clamps which was inevitable especially for the polyester composites reinforced by carbon or glass fibers. It is noteworthy that for the carbon fiber- and glass fiber-reinforced composites, the values of tan $\delta$ obtained by DMTA method are more than 2 or 3 times greater than those obtained from NDT method.

Repeatability of NDT method was also examined. As an illustration, only the obtained results for the polyester composite reinforced by glass fiber (i.e., PsGf specimen) are given in this work. The PsGf specimen was hit four times by wooden hammer and after each impulse, the produced sound was recorded and then analysed by FFT technique. Figure 10 shows the obtained results for four repetitions. As in Fig. 10, the first mode of vibration appears at frequency of approximately 3,800 Hz for each replicate. On the count of the fact that the vibration resonance frequency depends only on material nature, it seems reasonable that the first mode of vibration is detected at a relatively same frequency in each repetition of experiment. Values of elastic modulus and tan $\delta$ calculated for PsGf composite specimen after each replicate are shown in Fig. 11. These values clearly suggest that the NDT method produces reliable and precision data. It is noteworthy that the hammer impulses have no effect on the results obtained in several repetitions, because in this method, the viscoelastic properties could be calculated by obtaining resonance frequency, density, and damping energy that all depend on the nature of the test specimen.

## Conclusions

In this study, viscoelastic properties of three different polyester composites separately reinforced by carbon, glass, and hemp fibers as well as a polyester composite sample reinforced by poplar wood powder have been investigated through an NDT method based on longitudinal free vibration. Obtained results have been also compared with those provided from a destructive mechanical method (i.e., DMTA). Values of modulus of elasticity determined by NDT correspond to those obtained from DMTA, thus, an excellent agreement was discovered between the two different methods. Although a similar trend was found for loss parameter (tan $\delta$) measured by the different examined methods, the data obtained from DMTA approach are much greater than those calculated by means of NDT method that may be attributed to the sample preparation and/or the experimental procedure of the destructive dynamic mechanical method. However, the results reveal that the NDT method examined in this study is an accurate, reliable and easy-to-use approach to measure viscoelastic properties of polymeric composites. It should be noted that the above-mentioned equations can be exclusively used in longitudinal direction, but for flexural determination, new mathematical calculations are required and that will be presented in our future works.

**Acknowledgments** The authors are grateful to the Islamic Azad University-Science and Research Branch (IAUSRB) for financial support of the project.